\documentclass[aps,prb,twocolumn,superscriptaddress,showpacs]{revtex4-1}
\usepackage{graphicx,amsmath}

\begin{document}

\title{Nonadiabatic particle and energy pump at strong system-reservoir 
coupling}

\author{Eduardo C. Cuansing}
\email{eccuansing@up.edu.ph}
\affiliation{Institute of Mathematical Sciences and Physics, University of 
  the Philippines Los Ba\~{n}os, Laguna 4031, Philippines}
\author{Jian-Sheng Wang}
\email{phywjs@nus.edu.sg}
\affiliation{Department of Physics, National University of Singapore, 
  Singapore 117551, Republic of Singapore}
\author{Juzar Thingna}
\email{jythingna@ibs.re.kr}
\affiliation{Center for Theoretical Physics of Complex Systems, Institute 
  for Basic Science (IBS), Daejeon 34126, Republic of Korea}
\affiliation{Max-Planck-Institut f\"{u}r Physik komplexer Systeme, 
  N\"{o}thnitzer Str. 38, 01187 Dresden, Germany}

\date{March 10, 2020}

\begin{abstract}
We study the dynamics of electron and energy currents in a nonadiabatic 
pump. The pump is a quantum dot nanojunction with time-varying gate 
potential and tunnel couplings to the leads. The leads are unbiased and 
maintained at the same temperature and chemical potential. We find that 
synchronized variations of the gate and tunnel couplings can pump electrons 
and energy from the left to the right lead. Inspired by quantum heat 
engines, we devise a four-stroke operating protocol that can optimally pump 
energy and hence, we investigate energy transfer and the coefficient of 
performance of the device. We compare our device to a two-stroke pump and 
find that the latter's lower performance is due to the bi-directional flow 
of energy currents resulting in low net energy currents. The performance of 
our four-stroke pump can be improved, up to a point, by increasing the net 
energy carried by the pumped electrons through energy charging via the 
gate potential. This is achieved by increasing the durations of energy 
charging and discharging strokes in the pump's protocol. However, despite 
the large energy output for long charging and discharging strokes, the 
energy required to maintain the strokes become large too resulting in a 
stagnant pump performance. Our pump operates effectively only in the strong 
lead coupling regime and becomes a dud in the weak coupling regime due to 
the net output energy flowing in the reverse direction. We use 
nonequilibirum Green's functions techniques to calculate the currents and 
capture the effects of strong lead-channel coupling exactly while 
simultaneously incorporating three time-varying parameters. Results from 
our work could aid in the design of high-performance quantum pumps.
\end{abstract}


\maketitle

\section{Introduction}
\label{sec:intro}

Control of the flow of energy in nanoscale systems is important in an era 
where the generation of heat in high-speed electronic devices is a major 
problem. Phononics \cite{LiRMP12}, the study of the flow of energy carried 
by phonons in quantum systems, has led to the idea of quantum heat pumps 
that can transport energy even against a temperature bias \cite{PekolaPRL07}. 
In electronic transport, an analogous device called an electron pump, which 
can be a junction that uses gate potentials to propagate electrons without 
the action of a source-drain bias, has been studied extensively in both the 
adiabatic \cite{ThoulessPRB83,BrouwerPRB98,SwitkesSci99,EntinPRB02,
MoskaletsPRB02,BlumenthalNP07,RiwarPRB13,LudovicoPRB16} and the nonadiabatic 
\cite{StrassPRL05,BraunPRL08,MoskaletsPRB08,CavalierePRL09,CroyPRB12,
RocheNC13,KaestnerRPP15,HaughianPRB17,WenzAPL16,YamahataSR16} regimes. The 
pumping of energy carried by electrons has also been studied either in the 
weak system-environment coupling regime \cite{SegalPRE06} or in the adiabatic 
regime \cite{HumphreyPRL02,RenPRL10}, with the exception of specialized 
coherent transport models that have been studied in the non-adiabatic 
regime \cite{ReyPRB07}. 

Heat engines are thermodynamic devices that can transform input heat into 
work by following a specific operating protocol. Heat pumps, on the other
hand, require an input of work to propagate heat. Heat pumps are,
therefore, thermodynamic engines operated under a time-reverse 
protocol. It is thus natural to investigate the role of thermodynamic 
principles in enhancing the performance of heat pumps 
\cite{RenPRL10,PotaninaPRB19}. Inspired by heat engines that follow a
four-stroke protocol \cite{UzdinPRX15}, we design an electronic device that
can pump energy to a predefined direction in the absence of temperature
and voltage bias. The device operates in the non-adiabatic regime and is 
always strongly coupled to leads, in contrast to traditional adiabatic 
heat pumps that are weakly coupled to heat baths. In order to fully capture
the physics of this nontrivial operational regime, we employ nonequilibrium 
Green's functions techniques \cite{ArracheaPRB05,CroyPRB12,DarePRB16}. Our 
approach allows us to explore weak to strong system-environment coupling 
strengths and we find that even though we are able to pump unidirectional 
charge for any coupling strength, the unidirectional energy pumping occurs 
only in the strong coupling regime. In the nonadiabatic regime, we find 
values of the parameters wherein the output energy is maximum and 
accompanied by a coefficient of performance that is close to the maximum 
value too. Moreover, the device that we are proposing is feasible with 
current technology \cite{RocheNC13,KaestnerRPP15,HaughianPRB17,YamahataSR16} 
and hence, our results can be scrutinized experimentally.

Our paper is sectioned as follows. In Sec.~\ref{sec:model} we discuss the 
model of our proposed four-stroke non-adiabatic pump. We describe the four 
strokes of our device and contrast its performance with a minimal two-stroke 
pump. In Sec.~\ref{sec:current} we show how we calculate the electron and 
energy currents using nonequilibrium Green's functions in the time domain. 
In Sec.~\ref{sec:results} we present and discuss our results. The summary
and conclusion are in Sec.~\ref{sec:conclusion}.

\section{Modeling the Device}
\label{sec:model}

\begin{figure}[h!]
  \includegraphics[width=2.9in,clip]{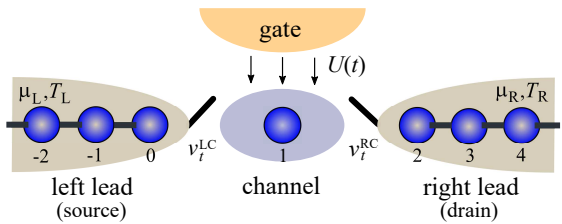}
  \caption{Electrons are pumped from the left lead to the right by 
  dynamically varying the tunnel couplings between the channel and the 
  leads, i.e., the $v_t^{\rm LC}$ and $v_t^{\rm RC}$, and the gate potential 
  $U(t)$ acting on the channel. The left and right leads have chemical 
  potentials $\mu_{\rm L}$ and $\mu_{\rm R}$ and temperatures $T_{\rm L}$ and 
  $T_{\rm R}$.}
  \label{fig:device}
\end{figure}

We consider a central one-site channel, tuned by a time-varying gate 
potential, coupled to left (the source) and right (the drain) leads, as 
illustrated in Fig.~\ref{fig:device}. The tunnel couplings between the 
channel and the leads can also be tuned dynamically and independently via 
additional gate potentials. In order to realize a pump, the variations of 
the gate potential and the tunnel couplings are synchronized according to 
a specific operating protocol (described below) so that electrons are 
unidirectionally pumped from the left to the right lead, even in the 
absence of a source-drain bias voltage. We model the device using the 
tight-binding approximation. The Hamiltonians for the leads are
\begin{equation}
  \begin{split}
    H^{\rm L} & = \sum_k \varepsilon_k^{\rm L} a_k^{\dagger} a_k + \sum_{k<j}
    v_{kj}^{\rm L} \left( a_k^{\dagger} a_j + a_j^{\dagger} a_k\right),\\
    H^{\rm R} & = \sum_k \varepsilon_k^{\rm R} b_k^{\dagger} b_k + \sum_{k<j}
    v_{kj}^{\rm R} \left( b_k^{\dagger} b_j + b_j^{\dagger} b_k\right),
  \end{split}
\end{equation}
where $a_k^{\dagger}$ ($b_k^{\dagger}$) and $a_k$ ($b_k$) are the spinless 
fermionic creation and annihilation operators at site $k$ in the left 
(right) lead, $\varepsilon_k^{\rm L}$ ($\varepsilon_k^{\rm R}$) are the on-site 
energies in the leads, and $v_{kj}^{\rm L}$ and $v_{kj}^{\rm R}$ are the 
nearest-neighbour hopping parameters. Sites in the left lead are labelled 
$k \in (-\infty, 0]$, while $k \in [2,\infty)$ are the labels of sites in 
the right lead. The single-site central channel Hamiltonian contains
\begin{align}
  H_0^{\rm C} = \varepsilon_1^{\rm C}c_1^{\dagger} c_1 \quad {\rm and} \quad
  H_t^{\rm C} = U\!(t)\,c_1^{\dagger} c_1,
\end{align}
where $c_1^{\dagger}$ and $c_1$ are the spinless fermionic creation and 
annihilation operators at site $1$, $\varepsilon_1^{\rm C}$ is the on-site 
energy at the same site, and $U\!(t)$ is the time-varying gate potential. 
The lead-channel coupling Hamiltonians are separated into stationary
\begin{equation}
  \begin{split}
    H_0^{\rm LC} & = v_{01}^{\rm LC}\left( a_0^{\dagger} c_1 + c_1^{\dagger} 
      a_0 \right),\\
    H_0^{\rm RC} & = v_{21}^{\rm RC}\left( b_2^{\dagger} c_1 + c_1^{\dagger} 
      b_2 \right),
  \end{split}
  \label{eq:Hcoup0}
\end{equation}
and time-dependent parts
\begin{equation}
  \begin{split}
    H_t^{\rm LC} & = v_{01}^{\rm LC}\!(t) \left( a_0^{\dagger} c_1 
      + c_1^{\dagger} a_0 \right),\\
    H_t^{\rm RC} & = v_{21}^{\rm RC}\!(t) \left( b_2^{\dagger} c_1 
      + c_1^{\dagger} b_2 \right).
  \end{split}
  \label{eq:Hcoupt}
\end{equation}
The time-varying $v_{01}^{\rm LC}\!(t)$ and $v_{21}^{\rm RC}\!(t)$ can increase 
or decrease the strength of the couplings between the channel and the leads. 
We set all of the hopping parameters to be space-symmetric, i.e., 
$v_{jk} = v_{kj}$. The total Hamiltonian is then the sum of the stationary 
and the time-varying parts, i.e., $H = H_0 + H_t$, where 
$H_0 = H^{\rm L} + H^{\rm R} + H_0^{\rm C} + H_0^{\rm LC} + H_0^{\rm RC}$ is the 
stationary part and $H_t = H_t^{\rm C} + H_t^{\rm LC} + H_t^{\rm RC}$ is the 
time-dependent part.

\begin{figure}[h!]
  \includegraphics[width=3.4in,clip]{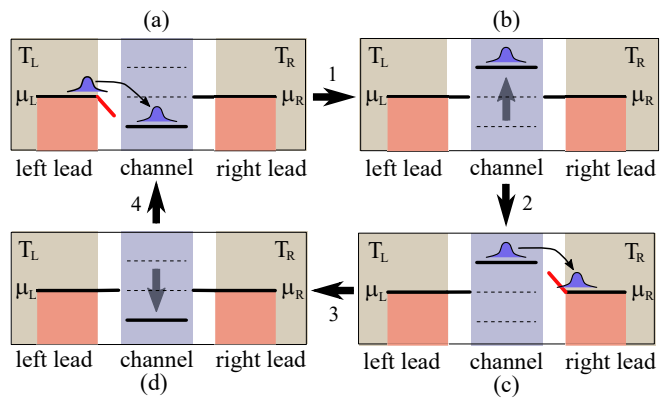}
  \caption{Illustration of the four strokes of the pump cycle. Electrons are 
  represented by (blue) wavepackets. (a) The left-channel coupling (in red) 
  allows more electrons from the left lead to flow into the channel. (b) The 
  left-channel coupling decreases and the gate potential in the channel is 
  raised thereby trapping some electrons within the channel. (c) The 
  right-channel coupling (in red) allows electrons to flow out to the right 
  lead. (d) The right-channel coupling decreases and the gate potential is 
  lowered thereby restricting the flow of electrons into or out of the 
  channel. The chemical potentials and temperatures of the two leads are the 
  same.}
  \label{fig:cycle}
\end{figure}

In order for the device to function as a pump, the time-dependence of the 
gate potential $U\!(t)$ and the tunnel couplings $v_{01}^{\rm LC}\!(t)$ and 
$v_{21}^{\rm RC}\!(t)$ must be synchronized. Inspired by four-stroke quantum 
heat engines such as the quantum Otto engine \cite{UzdinPRX15}, we design a 
four-stroke operating protocol as shown in Fig.~\ref{fig:cycle}. These four 
strokes are:
\begin{enumerate}
\item Stroke {\it a}: \emph{Transport stroke}. The channel and the left 
lead are coupled, the gate potential lowers the level in the channel, and 
electrons from the left lead can flow into the channel.
\item Stroke {\it b}: \emph{Energy charging stroke}. The left lead-channel 
coupling is abruptly decreased and, simultaneously, the gate potential 
raises the level in the channel. Electrons in the channel gain energy 
because of the raised level.
\item Stroke {\it c}: \emph{Transport stroke}. The channel and the right lead 
are coupled, the gate continues to raise the level in the channel, and 
electrons in the channel can flow to the right lead.
\item Stroke {\it d}: \emph{Energy discharging stroke}. The 
right lead-channel coupling is abruptly decreased and, simultaneously, 
the gate lowers the energy level in the channel.
\end{enumerate}
The cycle then repeats and at no point in time are the channel and leads 
disconnected. The driving potentials perform work on the device, pushing the
device into a nonequilibrium state which then allows us to pump electrons and 
energy despite the absence of temperature and voltage bias between the leads. 
There have been several studies that have investigated adiabatic pumping and 
the weak system-environment coupling regimes of quantum pumps 
\cite{ThoulessPRB83,BrouwerPRB98,ZhouPRL99,ThingnaPRB14}, with each attempt 
employing an approximate scheme to a specific physical time-dependent 
protocol. In the rest of this work, we will present an exact formulation to 
treat nonadiabatic pumping with three time-dependent components in a 
four-stroke operating protocol and contrast its performance with a minimal 
two-stroke protocol that comprises of strokes \emph{a} and \emph{c} only 
(see Fig.~\ref{fig:cycle}).

The energy gained by the pumped electrons in one cycle can be used to run 
another device such as a quantum motor \cite{FernandezPRB17} to convert 
electrical energy into mechanical energy. We can determine the energy 
current of the pumped electrons from the rate of change of the energy in 
each lead. The energy current \cite{CuansingPRB10,ThingnaPRB12,WangFP14,
EspositoPRL15}, traditionally known as the heat current, out of the left 
lead is
\begin{align}
  J^{\rm L}(t) = \left\langle -\frac{d H^{\rm L}}{dt}\right\rangle =
  -\frac{i}{\hbar} \left\langle \left[ H, H^{\rm L}\right]\right\rangle,
\end{align}
where the Heisenberg equation of motion is used in the second equality and 
considering that $H^{\rm L}$ has no explicit time dependence. The negative 
sign indicates that the current is positive if it is flowing to the right. 
The commutator between $H$ and $H^{\rm L}$ can be derived by using the 
fermionic anti-commutation rules, i.e., $\{ a_k^{\dagger},
a_j\} = \{ a_k, a_j^{\dagger}\} = \delta_{kj}$ and zero otherwise. Thus, we get
\begin{equation}
  \begin{split}
    J^{\rm L}(t) =\, & 2\,\varepsilon_0^{\rm L} \left[ v_{01}^{\rm LC} 
      + v_{01}^{\rm LC}\!(t)\right] {\rm Re}\!
    \left[ G_{10}^{{\rm CL},<}(t,t) \right]\\
    & + 2\,v^{\rm L}_{-1\,0} \left[ v^{\rm LC}_{01} 
      + v_{01}^{\rm LC}\!(t)\right] {\rm Re}\!\left[ 
      G^{{\rm CL},<}_{1\,-1}(t,t)\right],
  \end{split}
  \label{eq:JL}
\end{equation}
where the lesser channel left-lead (${\rm CL}$) nonequilibrium Green's 
function is defined as
\begin{align}
  G_{jk}^{{\rm CL},<}(t_1,t_2) = \frac{i}{\hbar}\left\langle 
    a_k^{\dagger}(t_2)\, c_j(t_1)\right\rangle
  \label{eq:GCLless}
\end{align}
and ${\rm Re}[\,]$ refers to the real part. The energy current flowing 
into the right lead, $J^{\rm R}(t) $, is similarly determined. The result has 
the same form as Eq.~(\ref{eq:JL}) but with the replacement of all the 
superscripts ${\rm L}$, the subscripts $0$ and $-1$, and the operator $a_k$ 
by the superscripts ${\rm R}$, subscripts $2$ and $3$, and the operator 
$b_k$, respectively. There is also an overall negative sign due to the 
reversed direction of the current in the definition. To be consistent with
$J^{\rm L}(t)$, a positive $J^{\rm R}(t)$ means a current that is moving to 
the right lead. The net energy current flowing across the device is 
$J_{\rm net}(t) = J^{\rm L}(t) + J^{\rm R}(t)$.

The electric current flowing out of the left lead follows the time rate of 
change of the number of electrons in the lead, where the number operator in 
the left lead is $N^{\rm L}=\sum_k a^{\dagger}_k a_k$,
\begin{equation}
  \begin{split}
    I^{\rm L}(t) & = \left\langle - q \frac{d N^{\rm L}}{dt}\right\rangle
    = -\frac{i q} {\hbar} \left\langle \left[ H,N^{\rm L}\right]\right\rangle\\
    & = 2 q \left[ v^{\rm LC}_{01} + v_{01}^{\rm LC}\!(t)\right] {\rm Re}\!\left[
      G^{{\rm CL},<}_{10}(t,t)\right],
  \end{split}
  \label{eq:IL}
\end{equation}
where $q$ is the electron charge. Notice that $I^{\rm L}(t)$ is proportional 
only to the first term of the energy current $J^{\rm L}(t)$ appearing in 
Eq.~\eqref{eq:JL} and hence, in general, the pumping of electrons does not
guarantee the pumping of energy, and vice versa. The electric current 
flowing into the right lead, $I^{\rm R}(t)$, is obtained using the same 
replacements described below Eq.~\eqref{eq:GCLless}. Consequently, the net 
electric current is $I_{\rm net}(t) = I^{\rm L}(t) + I^{\rm R}(t)$.

The coefficient of performance, ${\rm COP}$, of the pump can be determined 
from the ratio of the output energy and the net energy needed to run the 
device,
\begin{align}
  {\rm COP} = \frac{E_{\rm out}}{E_{\rm in}},
  \label{eq:COP}
\end{align}
where the output and input energies are
\begin{equation}
  \begin{split}
    E_{\rm out} & = \int_{\rm cycle} J_{\rm net}(t)\,dt,\\
    E_{\rm in} & = \frac{1}{T}\int_{\rm cycle}\left( \left|\langle H_t^{\rm C}\rangle
    \right| + \left|\langle H_t^{\rm LC}\rangle\right| + \left|\langle
    H_t^{\rm RC}\rangle\right|\right)dt.
    \label{eq:Einout}
  \end{split}
\end{equation}
The integrals are over one pumping cycle and $T$ is the cycle period. For
$E_{\rm in}$, which is always positive, we need the values of the energy that 
the time-varying gate supplies,
\begin{equation}
  \begin{split}
    \left|\langle H_t^{\rm C}\rangle\right| & = \left|\left\langle
    U(t)\, c_1^{\dagger}(t) c_1(t)\right\rangle\right|\\
    & = \hbar\, U(t) \sqrt{{\rm Re}\!\left[G_{11}^{{\rm CC},<}(t,t)\right]^2
      + {\rm Im}\!\left[G_{11}^{{\rm CC},<}(t,t)\right]^2},
  \end{split}
\end{equation}
where ${\rm Im}[\,]$ refers to the imaginary part and the Green's function
\begin{align}
  G_{11}^{{\rm CC},<}(t_1,t_2) = \frac{i}{\hbar}\left\langle c_1^{\dagger}(t_2)
  c_1(t_1)\right\rangle
\end{align}
and the energies supplied by the time-varying lead-channel couplings,
\begin{equation}
  \begin{split}
    \left|\langle H_t^{\rm LC}\rangle\right| & = \left| 2 \hbar\, v_t^{\rm LC}\,
    {\rm Im}\left[ G_{10}^{{\rm CL},<}(t,t)\right]\right|,\\
    \left|\langle H_t^{\rm RC}\rangle\right| & = \left| 2 \hbar\, v_t^{\rm RC}\,
    {\rm Im}\left[ G_{12}^{{\rm CR},<}(t,t)\right]\right|.
  \end{split}
\end{equation}
$E_{\rm in}$ is therefore the total amount of input energy per cycle needed 
to operate the device.

\section{Nonequilibrium Green's functions}
\label{sec:current}

\begin{figure}[h!]
  \includegraphics[width=0.9\columnwidth,clip]{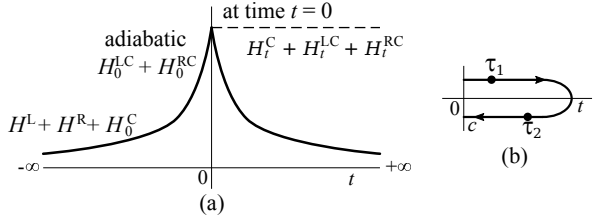}
  \caption{Illustrations of (a) how the various terms in the Hamiltonian are
    switched on and (b) the Keldysh contour $c$ with contour time variables 
    $\tau_1$ and $\tau_2$. The contour begins at $0$, goes to $t$, and then 
    back to $0$.}
  \label{fig:switch}
\end{figure}

We use the Schwinger-Keldysh formalism \cite{PastawskiPRB92,JauhoPRB94,
HaugSPR08,StefanucciCUP13} to determine the lesser nonequilibrium Green's 
functions. The ${\rm CL}$ contour-ordered Green's function is defined as
\begin{align}
  G_{jk}^{\rm CL}(\tau_1,\tau_2) = -\frac{i}{\hbar}\left\langle {\rm T}_c\,
  c_j(\tau_1) a_k^{\dagger}(\tau_2)\right\rangle,
  \label{eq:GCL}
\end{align}
where ${\rm T}_c$ is the contour-ordering operator along the Keldysh contour 
$c$ shown in Fig.~\ref{fig:switch}(b).

Terms in the total Hamiltonian are switched on according to the scheme shown 
in Fig.~\ref{fig:switch}(a). At time far in the past, the channel and the 
two leads are initially considered to be uncoupled and at their own 
equilibrium states. The lead-channel couplings are then adiabatically 
switched on in such a way that at time $t = 0$ the coupled system is at the 
steady state. The time-varying components 
$H_t = H^{\rm C}_t + H^{\rm LC}_t + H^{\rm RC}_t$ are then abruptly switched on at 
time $t = 0$. 

\begin{figure}[h!]
  \includegraphics[width=0.95\columnwidth,clip]{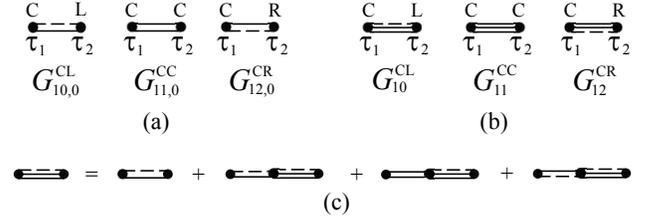}
  \caption{Diagram representations of (a) the steady-state Green's functions,
    (b) the nonequilibrium Green's functions, and (c) the approximate 
    iterative equation for $G_{10}^{\rm CL}(\tau_1,\tau_2)$. Each vertex implies 
    either a $U$, $v_t^{\rm LC}$, or $v_t^{\rm RC}$ coupling strength depending 
    on the Green's function immediately to the left of it. All internal 
    contour time variables are integrated out.}
  \label{fig:diagrams}
\end{figure}

In the interaction picture, the contour-ordered Green's function takes the 
form
\begin{align}
  G_{jk}^{\rm CL}(\tau_1,\tau_2) = -\frac{i}{\hbar} \left\langle
  {\rm T}_c\,e^{-\frac{i}{\hbar}\int_c H_t(\tau') d\tau'} c_j(\tau_1)\,
  a_k^{\dagger}(\tau_2)\right\rangle_0,
  \label{eq:GCLcontour}
\end{align}
where the subscript $0$ implies that the average is taken with respect to 
the steady state. Steady-state Green's functions can be determined exactly
(see below), even in the strong leads-channel coupling regime, because the 
associated stationary Hamiltonian $H_0$ is purely quadratic. The 
contour-ordered Green's function in Eq.~\eqref{eq:GCLcontour} is then 
determined via a diagrammatic perturbative expansion and the result is a 
series of terms containing steady-state Green's functions and their 
integrals. We would like to note that although the time-dependent 
perturbations $H^{\rm C}_t + H^{\rm LC}_t + H^{\rm RC}_t$ are quadratic in form, 
expanding the contour-ordered Green's function results in high-order 
diagrams that cannot be accounted by an iterative equation. In order to 
arrive at an iterative Dyson equation, we make an approximation by setting 
the amplitudes of the time-varying parameters $U\!(t)$, $v_{01}^{\rm LC}\!(t)$, 
and $v_{21}^{\rm RC}\!(t)$ to be much smaller than the on-site energy 
$\varepsilon_1$ and stationary hopping parameters $v_{01}^{\rm LC}$ and 
$v_{21}^{\rm RC}$. The diagram representations of the Green's functions are 
shown in Fig.~\ref{fig:diagrams}(a) and (b) and the resulting approximate 
iterative diagram equation is shown in Fig.~\ref{fig:diagrams}(c).

An alternative approach is to switch on the time-dependent perturbations at
a time when the channel and the leads are still uncoupled. In this case the
system is not in a steady state and the perturbation expansion of the
contour-ordered Green's function would lead to terms based on the 
equilibrium Green's functions of the leads and the channel. Although an 
exact iterative equation can be constructed from this approach, it requires 
knowledge of the channel's temperature and chemical potential, which are 
not well-defined due to the channel being finite. In contrast, in the 
approach that we use, there is no such requirement when the contour-ordered 
Green's function is expressed in terms of steady-state Green's functions.

Similar approximate iterative diagram equations can be derived for 
$G_{jk}^{\rm CR}$ and $G_{jk}^{\rm CC}$. The corresponding iterative equations 
for these contour-ordered Green's functions are
\begin{equation}
  \begin{split}
    G_{jk}^{\gamma}(\tau_1,\tau_2) = &~ G_{jk,0}^{\gamma}(\tau_1,\tau_2)\\
    &~ + \int_c d\tau'\, B_{j1,0}^{\rm C}(\tau_1,\tau')\,
    G_{1k}^{\gamma}(\tau',\tau_2),
    \label{eq:contour-ordered}
  \end{split}
\end{equation}
where $\gamma = {\rm CL}, {\rm CC}, {\rm CR}$, the subscript $0$ indicates 
the steady-state version of the Green's function, the integrals are along 
the Keldysh contour, and
\begin{equation}
  \begin{split}
    B_{j1,0}^{\rm C}(\tau,\tau') = &~ G_{j0,0}^{\rm CL}(\tau,\tau')\,
    v_{01}^{\rm LC}(\tau') + G_{j1,0}^{\rm CC}(\tau,\tau')\,U_1(\tau')\\
    & + G_{j2,0}^{\rm CR}(\tau,\tau')\,v_{21}^{\rm RC}(\tau').
    \label{eq:Bs}
  \end{split}
\end{equation}
Notice that only the steady-state versions of the Green's functions appear 
in Eq.~\eqref{eq:Bs}. Applying analytic continuation and Langreth's theorem
to the contour-ordered Green's function in Eq.~\eqref{eq:contour-ordered},
the retarded and advanced nonequilibrium Green's functions in real time
variables are
\begin{equation}
  \begin{split}
    G_{jk}^{\gamma,\alpha}(t_1,t_2) = &~ G_{jk,0}^{\gamma,\alpha}(t_1,t_2)\\
    &~ + \int_0^t dt'\, B_{j1,0}^{{\rm C},\alpha}(t_1,t')\,
    G_{1k}^{\gamma,\alpha}(t',t_2),
    \label{eq:Gra}
  \end{split}
\end{equation}
where $\alpha = r,a$. Also from the Langreth rules, the lesser nonequilibrium
Green's functions are
\begin{equation}
  \begin{split}
    G_{jk}^{\gamma,<}&(t_1,t_2) = G_{jk,0}^{\gamma,<}(t_1,t_2)\\
    + &~ \int_0^t dt'\, B_{j1}^{{\rm C},r}(t_1,t')\, G_{1k,0}^{\gamma,<}(t',t_2)\\
    + &~ \int_0^t dt'\, B_{j1,0}^{{\rm C},<}(t_1,t')\,
    G_{1k}^{\gamma,a}(t',t_2)\\
    + &~ \int_0^t dt' \int_0^t dt''\, B_{j1}^{{\rm C},r}(t_1,t')\,
    B_{11,0}^{{\rm C},<}(t',t'')\, G_{1k}^{\gamma,a}(t'',t_2),
    \label{eq:Gless}
  \end{split}
\end{equation}
where $\gamma$ is either ${\rm CL}$ or ${\rm CR}$ and
\begin{equation}
  \begin{split}
    B_{j1}^{{\rm C},\alpha}(t,t') = &~ G_{j0}^{{\rm CL},\alpha}(t,t')\,
    v_{01}^{\rm LC}(t') + G_{j1}^{{\rm CC},\alpha}(t,t')\, U_1(t')\\
    &~ + G_{j2}^{{\rm CR},\alpha}(t,t')\, v_{21}^{\rm RC}(t'),
    \label{eq:B}
  \end{split}
\end{equation}
while the $B_{j1,0}^{{\rm C},\alpha}(t,t')$ are the corresponding steady-state
versions. Note that the Green's functions in Eq.~\eqref{eq:Gless} are the
lesser nonequilibrium Green's functions needed to determine the currents.

To numerically determine the retarded and advanced nonequilibrium Green's 
functions in Eq.~(\ref{eq:Gra}), we discretize the time variable and 
re-express the integral as a sum \cite{CuansingIJMPB17}. Steady-state 
Green's functions are determined from the adiabatic switch-on of 
$H_0^{\rm LC} + H_0^{\rm RC}$, as shown in Fig.~\ref{fig:switch}(a), and leads
to an exact iterative Dyson equation. For the ${\rm CC}$ steady-state 
Green's function, we get
\begin{equation}
  \begin{split}
    G_{11,0}^{\rm CC}&(\tau_1,\tau_2) = g_{11}^{\rm C}(\tau_1,\tau_2)\\
    & + \int_c d\tau' \int_c d\tau''\,g_{11}^{\rm C}(\tau_1,\tau')\,
    \Sigma_{11}^{\rm C}(\tau',\tau'')\,G_{11,0}^{\rm CC}(\tau'',\tau_2),
  \end{split}
\end{equation}
where $g_{11}^{\rm C}$ is the equilibrium Green's function of the channel. The 
self-energy is
\begin{align}
  \Sigma_{11}^{\rm C}(\tau,\tau') = v_{10}^{\rm CL}\,g_{00}^{\rm L}(\tau,\tau')
  \,v_{01}^{\rm LC} + v_{12}^{\rm CR}\,g_{22}^{\rm R}(\tau,\tau')\,v_{21}^{\rm RC},
\end{align}
and $g_{00}^{\rm L}$ and $g_{22}^{\rm R}$ are the equilibrium Green's functions 
of site $0$ in the left lead and site $2$ in the right lead, respectively. 
Using analytic continuation and Langreth's theorem would lead to expressions 
for the retarded, advanced, and lesser ${\rm CC}$ steady-state Green's 
functions. Furthermore, since time-translation invariance is satisfied in 
the steady state, the steady-state Green's functions are simply functions 
of the difference between two times and we take their Fourier transforms 
into the energy domain to obtain
\begin{equation}
  \begin{split}
    G_{11,0}^{{\rm CC},r}(E) & = \left[ \left( E + i \eta\right)
      - \varepsilon_1^{\rm C} - \Sigma_{11}^{{\rm C},r}(E)\right]^{-1},\\
    G_{11,0}^{{\rm CC},a}(E) & = \left( G_{11,0}^{{\rm CC},r}(E)\right)^{\ast},\\
    G_{11,0}^{{\rm CC},<}(E) & = G_{11,0}^{{\rm CC},r}(E)\,\Sigma_{11}^{{\rm C},<}(E)\,
    G_{11,0}^{{\rm CC},a}(E).
  \end{split}
  \label{eq:GCC0}
\end{equation}
These ${\rm CC}$ steady-state Green's functions are often used in 
steady-state quantum transport calculations \cite{HaugSPR08,StefanucciCUP13}. 
Following the same procedure, the ${\rm CL}$ steady-state Green's functions 
are
\begin{equation}
  \begin{split}
    G_{jk,0}^{{\rm CL},r}(E) = &~ G_{j1,0}^{{\rm CC},r}(E)\,v_{10}^{\rm CL}\,
    g_{0k}^{{\rm L},r}(E),\\
    G_{jk,0}^{{\rm CL},a}(E) = &~ \left( G_{jk,0}^{{\rm CL},r}(E)\right)^{\ast},\\
    G_{jk,0}^{{\rm CL},<}(E) = &~ G_{j1,0}^{{\rm CC},r}(E)\,v_{10}^{\rm CL}\,
    g_{0k}^{{\rm L},<}(E)\\
    & ~~~+ G_{j1,0}^{{\rm CC},r}(E)\,\Sigma_{11}^{{\rm C},<}(E)\,
    G_{1k,0}^{{\rm CL},a}(E).
  \end{split}
  \label{eq:GCL0}
\end{equation}
Note that the Fourier transforms into the time domain of the steady-state
Green's functions are required in the calculation of the time-dependent
nonequilibrium Green's functions \cite{CuansingIJMPB17}.

The equilibrium Green's functions can be derived from the equation of
motion of the free leads. For the left lead, we find the retarded Green's
functions to be
\begin{equation}
  \begin{split}
    g_{00}^{{\rm L},r}(E) = &~ 2 \frac{\left( E + i \eta\right)
      - \varepsilon}{v^2} \pm 2\,i\,\frac{\sqrt{v^2
      - \left(\varepsilon-E\right)^2}}{v^2},\\
    g_{0\,-1}^{{\rm L},r}(E) = &~ \frac{2}{v^3}\left( 2\left( (E+i\eta)
    - \varepsilon\right)^2-v^2\right)\\
    & \pm i \frac{4}{v^3}\left((E-\varepsilon) \sqrt{v^2-(\varepsilon-E)^2}
    \right),
    \label{eq:gLr}
  \end{split}
\end{equation}
where we set all of the on-site energies to be
$\varepsilon=\varepsilon_j^{\rm L}$ and the hopping parameters to be
$v=v_{jk}^{\rm L}$ in the left lead. The advanced and lesser equilibrium 
Green's functions of the free left lead are
\begin{equation}
  \begin{split}
    g_{jk}^{{\rm L},a}(E) & = \left( g_{jk}^{{\rm L},r}(E)\right)^{\ast},\\
    g_{jk}^{{\rm L},<}(E) & = - f^{\rm L}(E)\,\left( g_{jk}^{{\rm L},r}(E)
    - g_{jk}^{{\rm L},a}(E)\right),
    \label{eq:gLal}
  \end{split}
\end{equation}
where $f^{\rm L}(E) = \left[\exp\left((E-\mu_{\rm L})/k_BT_{\rm L}\right)
+ 1\right]^{-1}$ is the Fermi-Dirac distribution containing information 
about the leads chemical potential and temperature. 

Expressions for the ${\rm CR}$ steady-state the and equilibrium Green's 
functions of the right lead can be similarly derived. The results are in the 
same form as Eqs.~\eqref{eq:GCL0}, \eqref{eq:gLr} and \eqref{eq:gLal} except 
for the replacement of all ${\rm L}$ superscripts with ${\rm R}$ and the 
corresponding site label $0$ subscripts with label $2$. In addition, the 
chemical potential and temperature of the right lead should be used in the 
Fermi-Dirac distribution.

\section{Results and Discussion}
\label{sec:results}

We determine the time-dependent electric currents, $I^{\rm L}(t)$ and 
$I^{\rm R}(t)$, and energy currents, $J^{\rm L}(t)$ and $J^{\rm R}(t)$, as
described in Sec.~\ref{sec:current}. These expressions require the 
calculation of nonequilibrium Green's functions, which are integrals of the 
steady-state Green's functions in the time domain. The steady-state Green's 
functions are determined in the energy domain and then Fourier transformed 
into the time domain. Both the Fourier transforms and the multiple 
integrals in the calculations of the Green's functions are numerically 
determined using standard numerical integration techniques \cite{PressCUP07}
while matrix manipulations are done numerically using LAPACK (Linear Algebra
Package) \cite{AndersonSIAM99}. We discretize the time using time steps of 
$0.1\,{\rm fs}$ and for every set of values of the parameters, we calculate 
the currents up to a total time of $5$ pumping cycles.

\begin{figure}[h!]
  \includegraphics[width=0.9\columnwidth,clip]{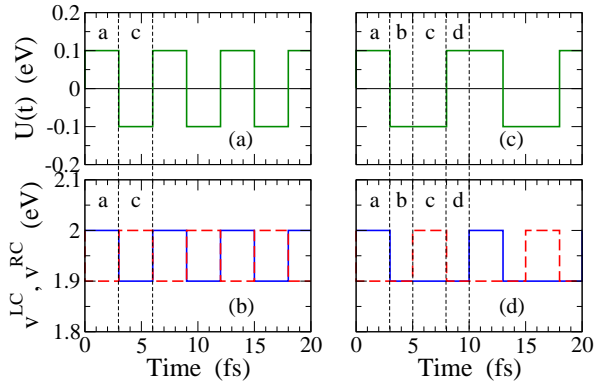}
  \caption{The protocols for the two-stroke pump ((a) and (b)) and the 
    four-stroke pump ((c) and (d)). Shown in plots (a) and (c) are the gate 
    potential $U(t)$ variations, by $0.1\,{\rm eV}$, while plots (b) and (d) 
    shows the time-dependence of the tunnel couplings 
    $v^{\rm LC}=v^{\rm LC}_{01}+v^{\rm LC}_{01}(t)$ (blue lines) and 
    $v^{\rm RC}=v^{\rm RC}_{21}+v^{\rm RC}_{21}(t)$ (red dashed lines), also by
    $0.1~{\rm eV}$. The durations of the strokes are indicated as vertical 
    dashed lines. Values of the other parameters we use are 
    $\varepsilon^{\rm L} = \varepsilon^{\rm R} = \varepsilon^{\rm C} = 1\,{\rm eV}$ 
    for the on-site energies, $v^{\rm L} = v^{\rm R} = v_{01}^{\rm LC} = 
    v_{21}^{\rm RC} = 2\,{\rm eV}$ for the static tunnel couplings, 
    $\mu_{\rm L} = \mu_{\rm R} = 0$ for the chemical potentials of the leads, 
    $E_f = 0$ for the Fermi energy, and $T_{\rm L} = T_{\rm R} = 300\,{\rm K}$ 
    for the temperatures of the leads.}
  \label{fig:protocols}
\end{figure}

We first study a minimal approach of pumping electrons and energy in our 
quantum dot device. The time-dependent protocol consists of only two 
strokes, {\it a} and {\it c}, as depicted in Fig.~\ref{fig:cycle}. In this 
case, we do not let the electrons charge or discharge via the action of the 
gate potential. Since only the transport strokes are involved, electrons 
would flow from the left lead to the channel in stroke {\it a} and then 
proceed to flow to the right lead in stroke {\it c}. Following this 
protocol, the pumped energy would be just enough to move the electrons 
across the device. The absence of the charging stroke does not allow the 
electrons to adjust to the higher gate potential within the channel. Thus, 
the electrons that are transported from the left lead to the channel in 
stroke {\it a} do not get a chance to fully gain energy within the channel, 
despite the increase in the gate potential. This is because the increase in 
the gate potential is accompanied by the opening of the transport channel to 
the right lead. Thus, even though we observe electron pumping in this case, 
the energy pumped during one cycle is not optimized and hence, there is no 
extra energy per cycle to be harvested from the pumped electrons.

\begin{figure}[h!]
  \includegraphics[width=0.95\columnwidth,clip]{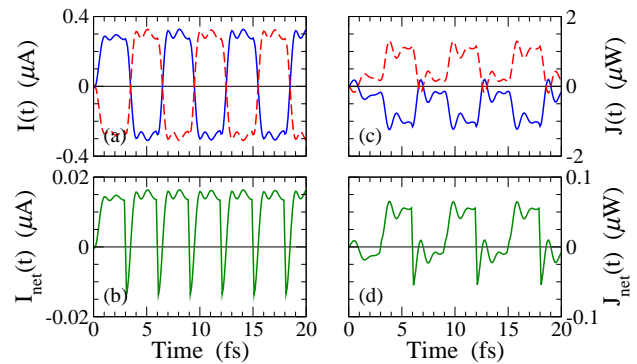}
  \caption{The currents as functions of time for the two-stroke pump. 
  (a) The current from the left lead $I^{\rm L}(t)$ (blue curve) and the 
  current into the right lead $I^{\rm R}(t)$ (red dashed curve). (b) The net 
  current $I_{\rm net}(t) = I^{\rm L}(t) + I^{\rm R}(t)$. (c) The energy current 
  from the left lead $J^{\rm L}(t)$ (blue curve) and the energy current into 
  the right lead $J^{\rm R}(t)$ (red dashed curve). (d) The net energy current
  $J_{\rm net}(t) = J^{\rm L}(t) + J^{\rm R}(t)$.}
  \label{fig:pumpcurrents}
\end{figure}

Our results are shown in Fig.~\ref{fig:pumpcurrents}. The gate potential 
and the lead-channel couplings are varied according to 
Figs.~\ref{fig:protocols}(a) and (b). The stroke durations are $3\,{\rm fs}$ 
each and the period of pumping cycle is $6\,{\rm fs}$. It is important to 
stress that even though our time-dependent protocols manipulate the transport 
through our nonadiabatic pump, the couplings to the leads are always 
non-zero, implying that electrons may always flow between the leads and 
the channel. As seen in Fig.~\ref{fig:pumpcurrents}(a) the electron currents 
alternate between flowing in and out of the channel. During stroke {\it a},
the gate potential lowers the energy level in the channel thus allowing
electrons to flow from the left lead to the channel. During stroke {\it c}, 
the gate potential now raises the energy level within the channel, while at
the same time the coupling strength to the left lead is decreased, thereby
allowing more electrons to flow from the channel to the right lead. In this 
minimal two-stroke protocol we get a perfect pumping of the electrons, as 
seen via the net current in Fig.~\ref{fig:pumpcurrents}(b). The net current 
is mostly positive indicating a flow of electrons from the left lead to the
right with sudden spikes reversing the current flow appearing at the 
transition points in-between the two strokes. These spikes could be reduced 
by smoothing out the time-dependent protocol instead of employing abrupt 
square wave pulses.

The energy current for the two-stroke pump tells a different story, as seen 
in Figs.~\ref{fig:pumpcurrents}(c) and (d). The left and right lead currents 
alternate in the same fashion as the electron current. However, the net 
energy current flows in the opposite direction to the electron current 
during stroke {\it a}. This implies that even though there are more 
electrons on average flowing to the right, the electrons flowing to the 
left have more energy hence reversing the net energy current during stroke 
{\it a}. This makes energy pumping in our two-stroke protocol non-ideal 
even though during one full cycle the net pumped energy is from left to 
right. For both electron and energy pumping, the transients last for only 
one cycle after which the device quickly approaches the periodic asymptotic 
state. The four-stroke non-adiabatic pump, discussed in Sec.~\ref{sec:model}, 
overcomes this drawback and causes even the energy current to flow from 
left to right throughout the cycle, except around sharp transitional 
points which can be prevented using a smooth protocol instead of an abrupt 
square wave.

\begin{figure}[h!]
  \includegraphics[width=0.95\columnwidth,clip]{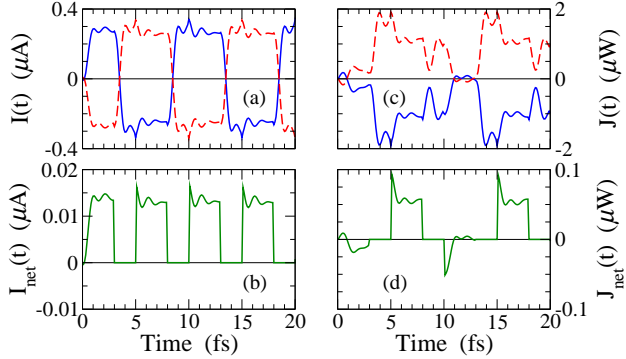}
  \caption{The pumped currents as functions of time for the four-stroke pump.
  Shown are (a) the electric currents $I^{\rm L}(t)$ (blue curve) and 
  $I^{\rm R}(t)$ (red dashed curve), (b) the net electric current 
  $I_{\rm net}(t)$, (c) the energy currents $J^{\rm L}(t)$ (blue curve) and
  $J^{\rm R}(t)$ (red dashed curve), and (d) the net energy current 
  $J_{\rm net}(t)$.}
  \label{fig:enhancedcurrents}
\end{figure}

In the four-stroke pump, we vary the gate potential and the tunnel couplings 
according to Figs.~\ref{fig:protocols}(c) and (d). The duration of the 
transport strokes {\it a} and {\it c} are $3\,{\rm fs}$ while the energy 
charging and discharging strokes are for $2\,{\rm fs}$. The period of the 
pumping cycle is $10\,{\rm fs}$ and stroke transitions are sudden and abrupt.
The pumped electron and energy currents are shown in 
Fig.~\ref{fig:enhancedcurrents}. As the pump is being operated, we see from
Fig.~\ref{fig:enhancedcurrents}(a) that the left and right currents are
alternating between flowing into and out of the channel, similar to the way
the currents flow in the two-stroke pump. The net electron current, however, 
is markedly different from that of the two-stroke pump. 
Fig.~\ref{fig:enhancedcurrents}(b) shows that the net electron current in 
each pumping cycle flows to the right. However, during energy charging 
and discharging strokes we see that the left and right currents exactly 
cancel, resulting in no net current flow. Similarly, the net energy 
current shown in Fig.~\ref{fig:enhancedcurrents}(d) shows no net energy 
current flowing during the energy charging and discharging strokes. 

\begin{figure}[h!]
  \includegraphics[width=0.98\columnwidth,clip]{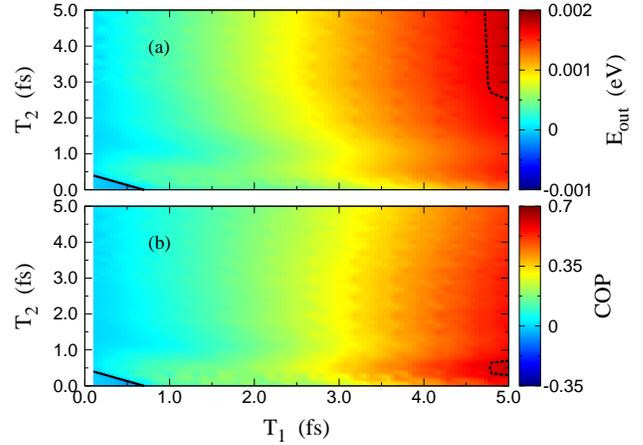}
  \caption{(a) The total output energy $E_{\rm out}$ per cycle and (b) the
  coefficient of performance ${\rm COP}$ per cycle when the transport
  stroke durations $T_1$ and charging and discharging strokes durations
  $T_2$ are varied. The line at the lower left corner of the plots 
  separates the regions of positive and negative $E_{\rm out}$ and ${\rm COP}$. 
  The dashed line encloses the regions where we obtain the maximum 
  $E_{\rm out}$ and ${\rm COP}$.}
  \label{fig:t1t2}
\end{figure}

The role of the charging and discharging strokes are to increase the energy
of the pumped electrons. Longer charging strokes, i.e., larger $T_2$, means 
more pumped energy $E_{\rm out}$ per pumping cycle. In terms of the pump's 
performance, however, longer charging strokes do not necessarily lead to 
better performance. In the ${\rm COP}$ defined in Eq.~\eqref{eq:COP}, the 
input energy $E_{\rm in}$, defined in Eq.~\eqref{eq:Einout}, also depends on 
the duration of the strokes. Shown in Fig.~\ref{fig:t1t2} are contour plots 
of the output energy $E_{\rm out}$ and the coefficient of performance 
${\rm COP}$ as the durations of the strokes are varied. $T_1$ is the duration 
of transport strokes {\it a} and {\it c} while $T_2$ is the duration of the 
energy charging stroke {\it b} and energy discharging stroke {\it d}. Note 
that there is no data for the $T_1=0$ line because we have a non-working 
pump when the transport strokes are off. Furthermore, the $T_2=0$ line 
indicates data for the two-stroke pump and shows the minimal values for 
$E_{\rm out}$ and the ${\rm COP}$. Notice that there is a region where both 
$E_{\rm out}$ and the ${\rm COP}$ are negative, indicating that energy is 
flowing in the opposite direction and we get a dud energy pump (see the 
lower left regions below the black lines in Fig.~\ref{fig:t1t2}). In this 
regime the stroke durations are too short and the system is continually in 
the transient regime where rapid oscillations occur after every abrupt 
stroke transition. The maximum ${\rm COP}$ and $E_{\rm out}$ appear in the 
region of large transport stroke durations $T_1$. Particularly, our pump 
operates with a relatively large energy output at high coefficient of 
performance (the regions enclosed by the dashed lines in 
Fig.~\ref{fig:t1t2}).

\begin{figure}[h!]
  \includegraphics[width=0.9\columnwidth,clip]{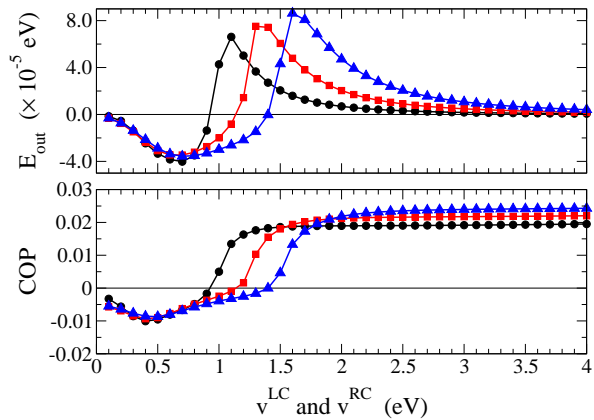}
  \caption{(a)The total output energy per cycle $E_{\rm out}$ and (b) the
  coefficient of performance per cycle ${\rm COP}$ as functions of the
  lead-channel couplings $v^{\rm LC}=v^{\rm RC}$. The values of the
  hopping parameters in the leads are $v^{\rm L} = v^{\rm R} = 2\,{\rm eV}$
  (circles), $2.5\,{\rm eV}$ (squares), and $3\,{\rm eV}$ (triangles). The
  on-site energies are maintained at $\varepsilon^{\rm L} = \varepsilon^{\rm R} 
  = 1\,{\rm eV}$. The amplitudes of the time-dependent perturbations are 
  $\Delta U=0.01\,{\rm eV}$ and $\Delta v^{\rm LC}=\Delta v^{\rm RC}=
  0.01\,{\rm eV}$. The durations of the strokes are $T_1=2\,{\rm fs}$ and 
  $T_2=3\,{\rm fs}$.}
  \label{fig:vCLdep}
\end{figure}

We have also investigated the effects of varying the lead-channel coupling 
on the performance of the pump. Shown in Fig.~\ref{fig:vCLdep} are the 
plots of $E_{\rm out}$ and the ${\rm COP}$ when the lead-channel couplings 
$v^{\rm LC}$ and $v^{\rm RC}$ are varied. In the regions where the couplings 
are weak, we find that $E_{\rm out}$ and the ${\rm COP}$ are negative 
indicating a dud energy pump, even though the electron current still flows 
from the left to the right lead. Transient oscillations after an abrupt 
stroke transition are large and long-lived when the couplings are weak. 
For $T_1 = 2\,{\rm fs}$ and $T_2 = 3\,{\rm fs}$, transient oscillations in 
energy have not dissipated enough resulting in a dud energy pump. As the 
coupling strengths become stronger, transient oscillations dissipate faster 
and an output energy that flows from the right to the left lead emerges. 
Notice that for a given value of $v^{\rm L}$ and $v^{\rm R}$ there is a, 
possibly resonance, value of around $v^{\rm LC}=v^{\rm L}/2$, and similary 
for $v^{\rm RC}$, where the $E_{\rm out}$ is maximum. At the strong coupling 
regions, we find that both $E_{\rm out}$ and the ${\rm COP}$ approach a 
constant value. Also in this region, higher hopping parameters $v^{\rm L}$ 
and $v^{\rm R}$ lead to slightly better ${\rm COP}$. This is because higher 
hopping parameters encourages the electrons to hop from site to site 
therefore resulting in better transport. In contrast, we have also 
investigated the effects of varying the on-site energies 
$\varepsilon^{\rm L}$, $\varepsilon^{\rm R}$, and $\varepsilon^{\rm C}$ as the 
couplings $v^{\rm LC}$ and $v^{\rm RC}$ are varied. Similar to 
Fig.~\ref{fig:vCLdep}, the $E_{\rm out}$ and ${\rm COP}$ approach a constant 
value as the couplings are increased. However, higher on-site energies 
result in slightly lower performance. This is because higher on-site 
energies encourages the electrons to stay within the site and is therefore 
detrimental to transport.

In this work, we have set the temperatures and chemical potentials of the 
leads to be the same, i.e., $T_{\rm L} = T_{\rm R} = T$ and 
$\mu_{\rm L} = \mu_{\rm R} = \mu$. The device, therefore, neither has a
temperature gradient nor a source-drain bias that can drive currents. The 
observed currents are due to the synchronized dynamics of the gate
potential in the channel and the tunnel couplings between the leads and
the channel. We notice, however, that the actual values of $T$ and 
$\mu$ do affect the amount of output energy $E_{\rm out}$ and the performance
${\rm COP}$ of the pump. Higher temperatures and chemical potentials
lead to increased $E_{\rm out}$ and ${\rm COP}$. This can be understood by 
noting that even though the temperatures and chemical potentials of the 
leads are the same, the outcome is the net flow of electrons from the left 
to the right lead. Electrons from a left lead with higher temperature or 
chemical potential will have more energy thereby resulting in an increase 
in the total output energy flowing to the right.

\section{SUMMARY AND CONCLUSION}
\label{sec:conclusion}

We model a pump using a nanojunction with time-varying tunnel couplings 
between the leads and the channel and a dynamic gate potential within the 
channel. We establish a two-stroke operating protocol and a four-stroke 
enhanced operating protocol for the pump. At least two transport strokes 
are needed to pump electrons from the left lead to the right. For the 
four-stroke pump, the two transport strokes are enhanced by an energy 
charging stroke and an energy discharging stroke so that the transported 
electrons gain extra energy when they reach the right lead.

We use nonequilibrium Green's functions techniques to calculate the 
electric and energy currents across the device. The technique allows us 
to establish strong coupling between the leads and the channel and also 
abrupt, nonadiabatic, changes in the gate potential and the tunnel 
couplings. A requirement that we employ so that we can arrive at an 
iterative Dyson equation is that the amplitude of the changes in the gate 
potential and the tunnel couplings are small compared to typical energy 
values, such as the on-site energies and the hopping parameters, in the 
model. Thus, our leads are always connected to the channel making the 
experimental realization feasible. We also calculate the total energy 
output to the right lead and the coefficient of performance per cycle of 
the pump.

Nonequilibrium Green's functions calculations show both left-moving and 
right-moving electrons and energy currents toward the left and right leads. 
The resulting net currents, however, indicate electric currents flowing 
from the left to the right lead, thereby pumping net electrons in 
this direction only. In contrast, we see that the dynamics of the energy 
current does not exactly follow that of the electron current. In instances 
where the stroke durations are short or the lead-channel couplings are 
weak, it is possible for the electric current to flow to the right while 
the energy current flows in the opposite direction. This happens when 
those electrons that flow to the left have more energy than those that 
flow to the right, even though there are more right-moving electrons. In
the four-stroke pump, the roles of the energy charging and discharging 
strokes are to enhance the pumped energy and improve the pump's performance.
Longer charging and discharging strokes result in an increase in the output
pumped energy. However, longer strokes do not necessarily lead to an
improved pump performance due to an accompanying higher input energy 
required to maintain those strokes.

\begin{acknowledgments}

We would like to thank Kicheon Kang, Horacio Pastawski, Sergej Flach, and 
Peter Talkner for insightful discussions. E. C. C. acknowledges suport from 
the ICTP Asian Network on Condensed Matter and Complex Systems. J. T. 
acknowledges support from the Institute for Basic Science in Korea 
(IBS-R024-Y2) and the Advanced Study Group (ASG) \lq\lq Open Quantum Systems 
far from Equilibrium\rq\rq at MPIPKS. J. S. W. acknowledges support from an 
MOE tier 2 grant number R-144-000-411-112.
\end{acknowledgments}

\bibliography{MyReferences}

\end{document}